\documentclass[aps,superscriptaddress,amsmath,amssymb,noshowpacs,twocolumn]{revtex4-1}
\usepackage{hyperref}
\usepackage{epsfig}
\usepackage{enumerate}
\usepackage{color}
\newcommand{\rem}[1]{}
\newcommand{\beq}{\begin{equation}}
\newcommand{\eeq}{\end{equation}}
\newcommand{\refe}[1]{(\ref{#1})}
\newcommand{\refE}[1]{Eq.~(\ref{#1})}
\newcommand{\reff}[1]{Fig.~(\ref{#1})}
\newcommand{\Ith}{I_{\rm th}} 
\newcommand{\Qs}{Q_{\rm s}} 
\newcommand{\Fg}{\mathfrak{F}}
\newcommand{\qav}[1]{\left\langle{#1 }\right\rangle}

\begin{document}

\title{Charge fluctuations in single-electron tunneling oscillations}

\author{C. Negri}

\author{F. Pistolesi}

% OLD AFFILIATION
%\affiliation{Laboratoire Ondes et Mati\`{e}re d'Aquitaine (UMR 5798), 
%Universit\'{e} de Bordeaux I and CNRS,
%351 Cours de la Lib\'eration, F-33405 Talence Cedex, France}
% NEW AFFILIATION 2011 ---> CHECK HERE
\affiliation{
Univ.\ Bordeaux, LOMA, UMR 5798, F-33400 Talence, France.\\
CNRS, LOMA, UMR 5798, F-33400 Talence, France.
}

\date{\today}

\begin{abstract}
It has been predicted that in the presence of a sufficiently high-dissipative environment transport 
in a small tunnel junction can become extremely regular, giving rise to the phenomenon
of single-electron tunneling oscillations.
Recent progress in detection of high-frequency current fluctuations 
and the interest in single-electron sources motivate further   
investigations on the expected accuracy of the charge oscillations 
as a function of the impedance of the environment.
In this paper we study theoretically the charge-fluctuation 
spectrum at finite frequency for the system at hand,
and investigate its behavior as a function of the external 
impedance. 
The evolution and the disappearance of the single-electron oscillations
peak is described by analytical and numerical methods.
\end{abstract}

\maketitle

%%%%%%%%%%%%%%%%%%%%%%%%%
% INTRODUCTION
%%%%%%%%%%%%%%%%%%%%%%%%%
\section{Introduction}

Coulomb blockade of voltage-biased tunnel junctions in the presence of a dissipative environment is a well-understood 
phenomenon of quantum transport (for a review see for instance Ref.~\cite{NazarovIngold}).
It has been shown that if the impedance of the environment $R$ is smaller than the 
tunnel junction resistance $R_t$, but larger than the quantum of resistance $R_Q=2\pi \hbar /{\rm e} \approx 25.8 k\Omega$
($\hbar$  being the reduced Planck constant and ${\rm e}$ the electron charge),
the current is suppressed for bias voltage $V$ smaller than the Coulomb gap ${\rm e}/2C$, associated with the 
capacitance $C$ of the junction ($C$ is typically in the fF range). 
This effect can be visible for temperatures lower than the Coulomb charging energy ${\rm e}^2/2C$ 
and for  $R_t \gg R_Q$, which assures suppression of cotunneling. 
In practice realizing a high-impedance environment with a flat frequency response till frequencies 
of the order of ${\rm e}^2/2C\hbar$ is a challenging experimental problem, 
as discussed in detail in Ref.~\cite{IngoldGrabertDevoret2}. 
Most experimental observations of Coulomb blockade phenomena in {\em single} tunnel junctions are actually 
done in an intermediate-impedance situation ($R\sim R_Q$), which leads at least to clearly non-linear characteristics of the junction; 
see for example Refs.~\cite{Popovic,Pierre}. 
Nevertheless, it is feasible to realize impedances of the order of some hundreds of k$\Omega$; for instance in the recent
Ref.~\cite{Hongisto} a resistance $R\sim0.4$M$\Omega$ has been realized.
For these values not only the suppression of the current at low-bias voltage should be visible, 
but also an appealing effect predicted in the 1980s \cite{GefenBenJacob,AverinLikharev,Likharev-newresults,Likharev-correlated}:
the single-electron tunneling oscillations (SETOs). 

The idea behind this effect can be understood in the simplest way in the limit 
$R, V \rightarrow \infty$ with $V/R=I_b$, so that the tunnel junction is current biased.
The current slowly charges the capacitance. When 
$V=Q/C$ reaches the threshold ${\rm e}/2C$ one electron
can cross the junction. If $I_b \ll {\rm e}/R_t C$
this will happen just after $V$ has reached the threshold.
The charge on the capacitance after the tunneling event will be $Q\simeq-{\rm e}/2$,
and $Q$ will start to slowly increase again. 
A time $\sim{\rm e}/I_b$ is needed before a new electron can cross
the junction and the sequence can start again.
The voltage at the junction will thus be periodically modulated
at a tunable frequency $\sim I_b/{\rm e}$.

Observation of this phenomenon is difficult. The implementation of the required strong-impedance environment 
has been realized by different authors using on-chip resistors 
\cite{Martinis,Kuzmin-usj,Cleland-charge,Cleland-env,PashkinKuzmin,Joyez,Zheng}.
An alternative approach has also been tried by designing the environment with tunnel-junction arrays
\cite{DelsingLikharev,DelsingClaeson}, exploiting the fact that a large number of arrays 
reduces the stochastic nature of the current \cite{LikharevBakhvalov}. 
Recently, observation of soliton-like single-electron oscillations with this method has been reported \cite{Bylander}.
A similar phenomenon for superconducting Josephson junctions
has been predicted (Bloch oscillations) \cite{LikharevZorin} and investigated 
by many authors \cite{KuzminHaviland,Kuzmin,Corlevi2,Hekking}.
Arrays of dc SQUIDs have also been exploited in this case to build up the proper environment to obtain Coulomb blockade of Cooper pairs 
\cite{WatanabeHaviland,WatanabeHaviland2,Corlevi}.
Reports on the observation of Shapiro-step-like structures in microwave-irradiated 
junctions constitute the present state of the art for the experimental probe of this effect \cite{Kuzmin,PashkinKuzmin,MaibaumZorin}. %CHECK CITATION 

Progress in the detection of high-frequency current fluctuations \cite{AguadoKouw,Glattli,Reulet,Deblock}
%check in the references of these groups for other eventual interesting citations 
can open new possibilities of observation of this phenomenon and 
of the crossover region, where the oscillations are not completely established.
At the same time the possibility of generating a periodic and frequency-tunable 
electric signal without any oscillating source is an interesting opportunity and could have 
applications, for instance, as a motion actuator in nanomechanical systems or a controlled 
single-electron source. 

It should be mentioned that Coulomb blockade is instead easily observed in {\em double} tunnel junctions even in absence of an environment. 
This happens since the second tunnel junction plays the role of the large-impedance 
environment suppressing quantum charge fluctuations and preventing 
tunneling if the voltage is below a threshold.
%
%Note however that the threshold is different from ${\rm e}/2C$, since charge 
%fluctuations between the two tunnel junctions is possible, and the 
%final state has to be found minimizing the total electrostatic
%energy of the two capacitances \cite{NazarovIngold,OdintsovGrabert}.
%
Note however that this configuration (without environment) will not give rise to single-electron 
oscillations, but to the well known sequential transport regime.
Electrons hop stochastically through the first and then the second 
junctions: The advantage of an ohmic environment is that it 
can generate (at least in principle) a stable current source.

In this paper we study how accurate the SETOs can be as a function of the 
impedance and of the bias conditions. 
In order to do this we will study the charge-fluctuation spectrum 
at the junction capacitance (or equivalently the current-fluctuation 
spectrum through the resistance load), 
and in particular the width of the peak at the frequency
$\sim I_b/{\rm e}$.
Notwithstanding the relatively large number of papers on this subject, 
some specifically addressing the dynamics of arbitrarily biased mesoscopic tunnel junctions 
with an analytical approach \cite{Ueda,UedaYamamoto,UedaHatakenaka},
a consistent calculation of these quantities is not available. 
It can be useful in order to evaluate the expected effect in view of measuring current 
fluctuations in this kind of device. 
In this paper we show that the width of the peak scales as the inverse of the 
impedance for large $R/R_t$.
The peak remains observable till values of $R/R_t$ of the order 
of 5.

The plan of the paper is the following.
In Sec.~\ref{model} we present the model 
describing the transport through the 
junction.
In Sec.~\ref{rregimes} we discuss the different 
regimes that the system undergoes by varying the 
bias voltage and the environment resistance.
In particular we obtain an analytical expression for 
the the $I$-$V$ characteristic in the SETOs regime.
In Sec.~\ref{SQQ} we calculate analytically 
the charge spectrum. The results are
discussed and compared with numerical
Monte Carlo simulations in Sec.~\ref{MCresults}.
Section \ref{conclusions} gives our conclusions.

%%%%%%%%%%%%%%%%%%%%%%%%%%%%
% THE MODEL
%%%%%%%%%%%%%%%%%%%%%%%%%%%%

\section{The system and the model}
\label{model}

\begin{figure}\centering
\includegraphics[width=.4\textwidth]{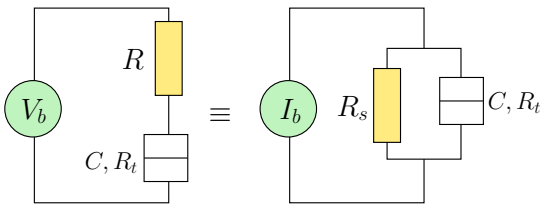}
\caption{\label{Model:fig} 
Circuit scheme of the system considered: a tunnel junction with capacitance 
$C$ and tunneling resistance $R_t$, biased by the constant current $I_b$ 
and shunted by a resistance $R_s$ in parallel (right), which is equivalent to a 
junction biased by a voltage $V_b$ and with a resistance $R$ in series (left) 
provided that $V_b=I_bR$ and $R=R_s$.
}
\end{figure}

Let us consider a tunnel junction 
with tunneling resistance $R_t \gg R_Q$ 
and associated capacitance $C$.
The circuit is voltage biased (at voltage $V_b$)
at zero temperature ($T=0$) in the presence of a resistor of 
resistance $R$ in series with the junction (see \reff{Model:fig}, left side).
This circuit is equivalent to one with a current source $I_b$ and a shunt resistor $R_s$ in parallel to the junction, 
provided $I_b=V_b/R$ and $R=R_s$ (see \reff{Model:fig}, right side).
We will thus use the parallel configuration to describe the device
in analogy with the previous literature \cite{AverinLikharev}. %, see \reff{Model:fig}.
It is clear that the results can be readily converted to the voltage 
bias case.
In particular note that the limit $R_s\rightarrow \infty$ (specifically $R_s \gg R_t$) 
describes the ideal current source.

%%%%%%%%%%%%%%%%%%%%%%%%%%%%%%%%%%%%%%%%%%%
%
%%%%%%%%%%%%%%%%%%%%%%%%%%%%%%%%%%%%%%%%%%%%%

Since we are interested in studying SETOs we need $R_s \gtrsim  R_t \gg R_Q$.
We thus assume from the outset that $R_s \gg R_Q$, which allows to neglect
quantum fluctuations and treat the charge degrees of freedom classically \cite{AverinLikharev}.
We also assume that the environment has a flat frequency response
$Z(\omega)=R_s$ up to frequencies $\hbar \omega \approx {\rm e}^2/2C$.
This hypothesis, though not easy to fulfill in practice, is the 
common assumption in the literature about this problem, and allows a 
simpler and more transparent approach.
In this regime transport through the junction is described by 
the theory currently known as orthodox Coulomb blockade theory.
Specifically the electron-tunneling rate depends on the 
voltage at the junction ($V_J$) as follows:
\beq
	\Gamma(V_J)= \theta(V_J-{\rm e}/2C) (V_J-{\rm e}/2C)/({\rm e}R_t)
	\label{rate}
\,,
\eeq
where $\theta$ is the Heaviside function.
[At finite temperature the function $\theta(V)$ is 
substituted by $1/(e^{-{\rm e}V/k_B T}-1)$.]
If $(R_Q\!\ll )\; R_s \!\ll \!R_t$ the standard picture of Coulomb blockade applies to the 
degrees of freedom of the environment:
They have the time to relax to thermal equilibrium 
between two electron-tunneling events.
The current-voltage characteristic in this case is then given by 
$I_J(V_J)=(V_J-{\rm e}/2C)\theta(V_J-{\rm e}/2C)$, exposing a clear Coulomb gap
for the current $I_J$ through the junction.
But if $R_s \sim R_t$ or larger (for moderate values of the 
bias voltage) the resistive environment (described for instance by 
a large collection of bosonic modes) reaches thermal equilibrium,
but not the charge on the capacitance $Q(t)$, which needs 
a time $\tau_s=R_s C$ to relax to its stationary state.
Formula \refe{rate} still holds, but with a time-dependent voltage $V(t)=Q(t)/C$.
The time dependence of the charge is given by
the solution of the differential equation:
\beq
	\dot Q = -Q/R_sC + I_b
	\label{circuit}
\,,
\eeq
which for an initial condition $Q_0$ at $t=0$ reads
\beq
	Q_f(Q_0,t)=(Q_0-I_b \tau_s) e^{-t/\tau_s}+I_b \tau_s
	\label{Qfunction}
\,.
\eeq
The stochastic problem is then completely formulated and 
in the remainder of the paper we discuss the behavior of the 
current and of the charge as a function of the two relevant 
dimensionless parameters of the problem: $\rho=R_s/R_t$
and $\kappa=(I_b-\Ith)/\Ith $, with $\Ith={\rm e}/2\tau_s$ the 
threshold for the current to start flowing through the 
tunnel junction.

%%%%%%%%%%%%%%%%%%%%%%%%%%%%%%%%%%%%%%%%%%%%%%%
%
\begin{figure}\centering
\includegraphics[width=.4\textwidth]{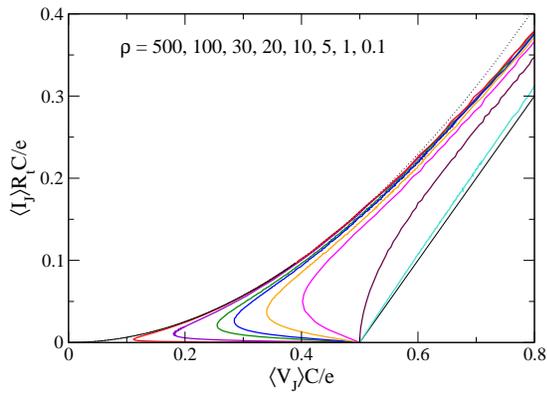}
\caption{\label{IjVj:fig} 
(Color online) Average current through the junction versus average voltage for different values of the ratio $\rho=R_s/R_t$.
The curves evolve from the standard Coulomb blockade suppression (rightmost line corresponding to $\rho=0.1$) 
to a square root behavior (leftmost line $\rho=500$). The dotted line is the large-$\rho$ limit of Eq.~\eqref{SETOextreme}.
}
\end{figure}
%
%
%%%%%%%%%%%%%%%%%%%%%%%%%%%%%%%%%%%%%%%%%%%%%

The current through the junction has already been calculated numerically 
in the very early literature \cite{Likharev-newresults}. 
We obtained the same results by Monte Carlo simulation
and, for convenience, we reproduce the curves in \reff{IjVj:fig}.
The limit of infinite $R_s$, or ideal current source, was discussed 
in details in Refs.~\cite{AverinLikharev,Zaikin}.
There it was shown in particular that in this limit
the system is in the SETOs regime with frequency $I_b/{\rm e}$ 
with an averaged voltage at the junction given by (see \cite{AverinLikharev}): 
\beq
	\langle V_{\rm J}\rangle=\sqrt{{\pi {\rm e} R_t\langle I_{\rm J}\rangle} \over {2C}}%={{\rm e} \over 2C} \sqrt{\pi(\kappa+1)\over \rho}
	\label{SETOextreme}
	\,.
\eeq
In the next section we will discuss the behavior of the system
for the intermediate regimes appearing when $R_s/R_t$ is not 
infinite, deriving in particular new analytical expressions for the 
SETOs frequency and $I$-$V$ characteristics.

%%%%%%%%%%%%%%%%%%%%%%%%%%%%
% REGIMES
%%%%%%%%%%%%%%%%%%%%%%%%%%%%

\section{Regimes of current transport}
\label{rregimes}

\begin{figure}
\includegraphics[width=.45\textwidth]{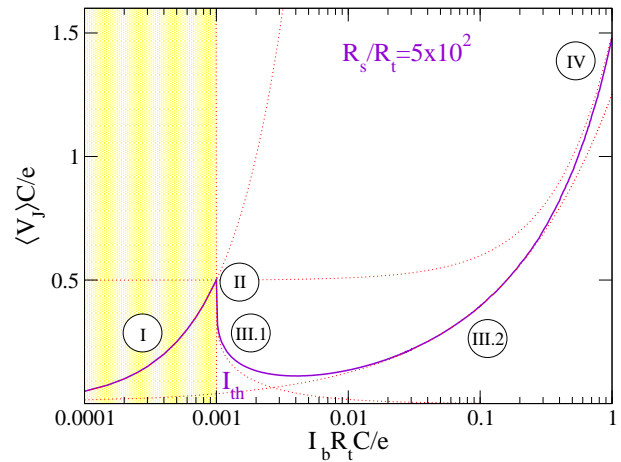}
\caption{
(Color online) Average voltage through the junction versus bias current in logarithmic scale ($\rho=5\times10^2$): 
Four different transport regimes can be outlined. 
The solid line gives the results of the numerical simulations, while dashed lines represent the analytical 
curves given in Sec.~\ref{rregimes}.
}
\label{regimes:fig} 
\end{figure}

\begin{figure}
\includegraphics[width=.45\textwidth]{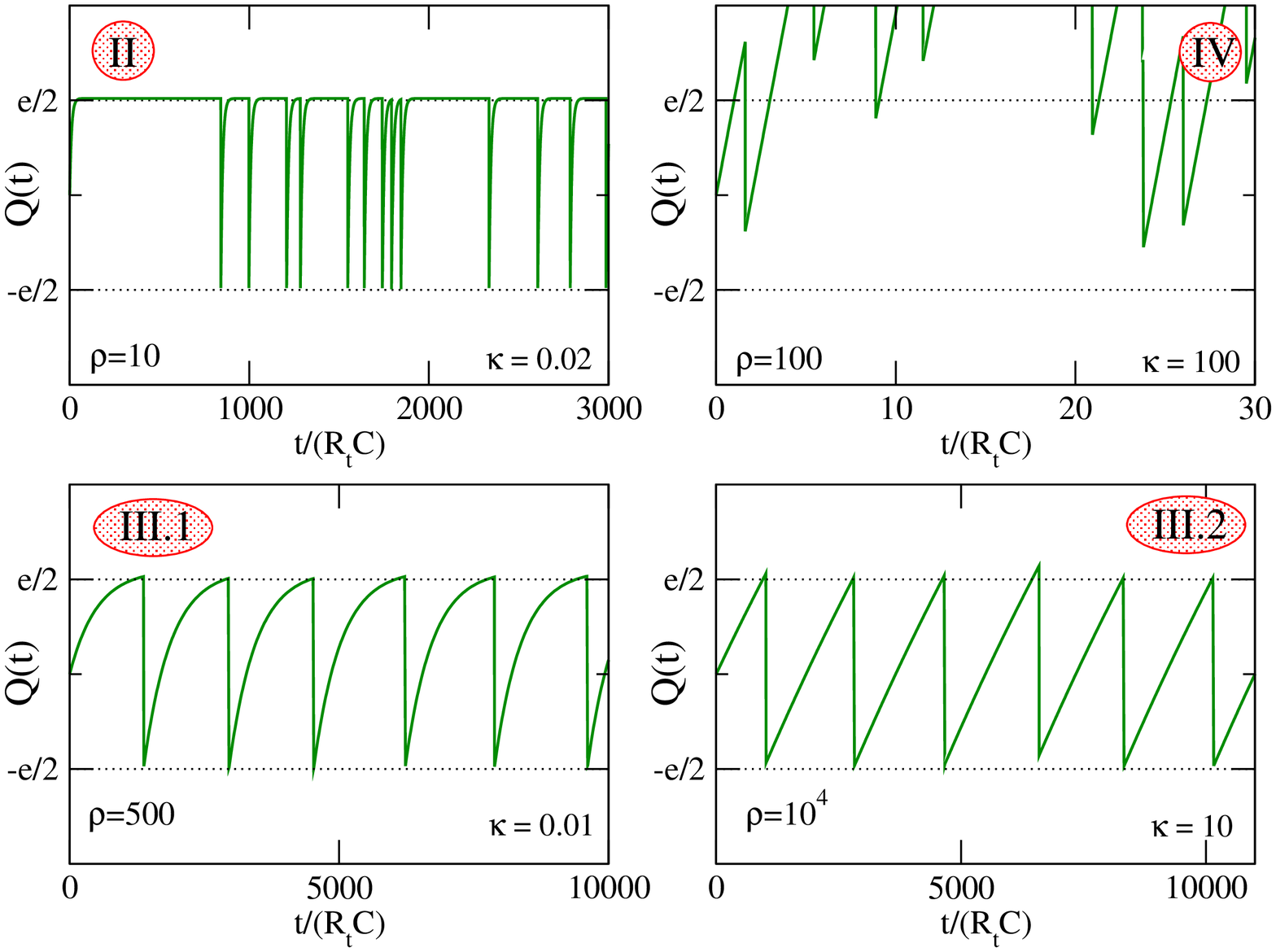}
\caption{
(Color online) The time behavior of the charge in the different regimes. 
Extreme values of the parameters $\rho$ and $\kappa$ have been 
chosen to best outline the differences.
}
\label{visualcharges:fig} 
\end{figure}

In this section we study the evolution of the current through the junction as a function 
of the current bias for different values of the external resistance.
The most interesting case is when $R_s/R_t$ is very large, we thus plot 
in Fig.~\ref{regimes:fig} the current on a logarithmic scale for the extreme 
value of $R_s/R_t=5 \times 10^2$.
\reff{regimes:fig} will be used as a ``map'' for the rest of the section,
where we will discuss how the junction evolves through the four different regimes
of transport indicated in the figure with roman numerals from (I) to (IV).
\reff{visualcharges:fig} shows the behavior of $Q(t)$ in the different regimes.

%%%%%%%%%%%%%%%%%%%%%%%%%%%%%%%%%%%
%
%%%%%%%%%%%%%%%%%%%%%%%%%%%%%%%%%%%
%
%

%%%%%%%%%%%%%%%%%%%%%%%%%%%%%%%%%%
%
%%%%%%%%%%%%%%%%%%%%%%%%%%%%%%%%%%%
%
%
\subsection{Non SETOs regimes}

Let us begin with the region indicated with (I) in 
\reff{regimes:fig}. 
For $I_b<\Ith$ the whole current 
flows through the shunting resistance and the voltage 
at the junction 
\beq
	\langle V_{\rm J}\rangle_I=R_s I_b
\eeq
remains below the threshold of the Coulomb blockade: ${\rm e}/2C$. 
Note that this first branch of the $I$-$V$ characteristic 
in \reff{IjVj:fig}  is flattened on the $I_J=0$ value 
and it is thus not visible.

For $I_b>\Ith$ transport through the junction becomes possible. 
For $I_b-\Ith \rightarrow 0^+$ one can identify  
a Poissonian regime of transport (region (II) in \reff{regimes:fig}), 
where the time between tunneling events fluctuates strongly. 
This is due to the fact that $\Gamma(V)$ as given by 
\refE{rate} vanishes linearly
near the threshold, and for very small $I_b-\Ith$ the charge has 
always the time (typically $\tau_s$) to reach the saturation value 
$\Qs=I_b \tau_s$. 
The typical inverse time between two tunnel events is: 
\beq
	1/\tau_{\rm eff} = \Gamma(\Qs)=(R_s/R_t) (I_b -I_{\rm th})/{\rm e} \ll 1/\tau_s \, .
\eeq
The last inequality sets also the region of existence of the regime (II) 
i.e.~, $0<\kappa \ll 1/\rho$.
%
%$I_b>I_{\rm th}$ and $(I_b-I_{\rm th})/I_{\rm th} \ll R_t/R_s$.
%
The average $V_J$ can be readily evaluated
by averaging the oscillations of the charge 
$Q(t)=\Qs-{\rm e}\,e^{-t/\tau_s}$
on the average time between two tunneling events 
$\tau_{\rm eff}+\tau_s \approx \tau_{\rm eff}$.
This gives:
\beq
	\label{II}
	\langle V_{\rm J}\rangle_{II}=R_s \left[ I_b(1-R_s/R_t)+I_{\rm th} R_s/R_t\right]
\eeq
and the curve is shown dashed in \reff{regimes:fig}.
Note that the slope changes sign at $R_s/R_t=1$.
We will
see that for $R_s\ll R_t$ this region joins continuously region (IV)
without the appearance of region (III).

It is thus convenient to discuss now the region (IV) 
defined as the limit of large $I_b$.
In this limit the junction has a $I$-$V$ characteristic of 
a normal resistor shifted by the Coulomb gap.
The average $V_J$ reads then:
\begin{equation}\label{IV}
        \langle V_{\rm J}\rangle_{IV} 
	=\frac{R_sR_t}{R_s+R_t}\left(I_b+\frac{{\rm e}}{2R_tC}\right) 
	\;.
\end{equation}
This expression holds for $\langle V_{\rm J} \rangle  \gg {\rm e}/2C$, 
i.e.\ for $I_b/\Ith \gg (R_t+R_s)/R_s$.
For $R_s \gg R_t$ it is then clear that a large region defined by the condition
\beq
%	{R_t}/{R_s} \ll {(I_b-I_{\rm th})}/{I_{\rm th}} \ll {R_s}/{R_t}
	1/\rho \ll \kappa \ll \rho
	\label{conditionSETO}
\eeq
exists between region (II) and region (IV). 
This is the SETOs region, (III) in \reff{regimes:fig}, which will be discussed below.
On the other side, for $R_s\ll R_t$, one sees that region (II) and region (IV) 
overlap at $ \kappa \approx 1$.
Actually it is straightforward to check that 
\refE{IV} expanded to first order in $R_s/R_t$ coincides with \refE{II}.

\subsection{SETOs regime}

Let us now discuss the single-electron oscillations regime, defined 
as region (III) in \reff{regimes:fig}.
This region is present only if $R_s\gg R_t$ and is characterized
by nearly periodic electron tunneling events,
since the time between two events is dominated by the 
deterministic charging time of the capacitance.
This time is typically of the order of $t_\star$, defined as  
the time needed to charge the capacitance from $Q=-{\rm e}/2$ to $Q={\rm e}/2$:
\beq
	\label{tstaradi}
      t_\star/\tau_s=
	\ln\left(\frac{I_b +I_{th} }{I_b -I_{th}}\right) =\ln {\left(2+\kappa \over \kappa\right)}
	\,.
\eeq
The electrons hop just after the threshold voltage has been reached. 

A general statistical theoretical framework is presented in 
Ref.~\cite{AverinLikharev} (and recalled in the Appendix),
but its analytical solution is given there only in the ideal current 
source limit ($R_s/R_t \rightarrow \infty$).
Actually in the case of the SETOs the approach is simplified and 
further progress is possible.
In order to obtain these results and for the calculation of the 
correlation function of the next section it is convenient to 
introduce a few concepts.
%
%%%%%%%%%%%%%%%%%%%%%%%%%%%%%%%%%%
%
%       FIGURE 4
%
%%%%%%%%%%%%%%%%%%%%%%%%%%%%%%%%%%
%
%
\begin{figure}\centering
%\vspace{0.5cm}
\includegraphics[width=.45\textwidth]{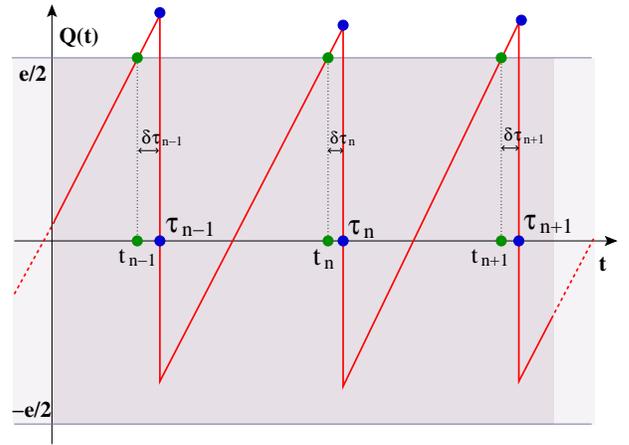}
\caption{\label{modeltimestep} 
(Color online) Details of the tunneling process: $t_n$ is the time at which the charge reaches the blockade region border after the $(n-1)$th tunneling \
event and $\delta\tau_n$ is the time it stays outside the border before the $n$th event. 
}
\end{figure}
%%%%%%%%%%%%%%%%%%%%%%%%%%%%%%%%%%%%%%%%%
%
%
In \reff{modeltimestep} the typical time dependence 
of the charge $Q(t)$ in the SETOs regime is shown. 
We can associate a number $n$ to each hopping 
event and define $t_n$ and $\tau_n=t_n+\delta \tau_n$ as the 
instant of time when $Q(t)={\rm e}/2$ and when the hopping 
event takes places, respectively (see \reff{modeltimestep}).
These quantities fluctuate randomly, but a correlation
between $t_n$ and $t_{n-1}$ exists.
Inversion of \refE{Qfunction} gives the time needed to reach the 
border of the Coulomb blockade region starting from a 
charge $Q_0$:
\begin{equation}
	\label{xi}
	\Xi(Q_0)=-\tau_s \ln\left(\frac{I_b\tau_s-{\rm e}/2}{I_b\tau_s-Q_0}\right)\,.
\end{equation}
The following relation between successive times $t_n$ holds:
\begin{equation}\label{Fgothdef}
	 t_n-t_{n-1}\equiv\mathfrak{F}(\delta\tau_{n-1})\,,
\end{equation}
with
\begin{eqnarray}
	\label{Fgoth}
 	\mathfrak{F}(\delta\tau)
	&=& \delta\tau+\Xi\left(Q_f\left(\frac{{\rm e}}{2},			 
	 \delta\tau\right)-{\rm e}\right) \nonumber \\
	&=&
	\delta \tau-\tau_s \ln\left({\kappa \over 2 + \kappa e^{-\delta \tau/\tau_s}}\right)
\,.
\end{eqnarray}

Let us now introduce the probability $P_n(t)$ that $n$ electrons have tunneled 
through the junction at the time $t$.
Within the SETOs region, this quantity is different from zero 
only in a small time region of the order of $t_\star$, and, in particular,
the above mentioned condition on the typical hopping time
implies that $P_n(t)$ will reach $1$ and then 
vanish in a time much shorter than $t_\star$, just after $Q(t)$ crosses the threshold ${\rm e}/2$. 
The rate equation for $t\geq t_n$ [$Q(t_n)={\rm e}/2$] takes the simple form:
\beq
	{d P_n\over dt}(t) = -\Gamma(Q_f({\rm e}/2,t)/C) P_n(t)
\eeq
with the initial condition $P_n(t_n)=1$.
The solution reads:
\beq
	P_n(t)=e^{%\exp 
	%\left\{
      -\frac{(I_b \tau_s-{\rm e}/2)}{{\rm e} R_t C}\tau_s \left(\frac{t-t_n}{\tau_s}+
	\emph{e}^{-{t-t_n \over \tau_s}}-1\right) }
	%\right\}
	\label{rateEq}
\,.
\eeq
For short times ($ t-t_n \ll \tau_s$) it has a Gaussian form 
\beq
	P_n(t) \approx \exp\left\{ -(t-t_n)^2\kappa  \rho/(4 \tau_s^2)  \right\}
	\label{sigmaGaussian}
\eeq  
with a decay time scale 
$
  \sim \tau_s /\sqrt{\kappa\rho} \ll t_\star
$ 
in  region (III).
The Gaussian form will thus be used in the following for the analytical
calculations.

From the knowledge of $P_n(t)$ it is possible to
obtain the probability density that a hopping event takes places 
at time $t$: ${\cal P}(t) = - d P_n/dt$, for $t\geq t_n$.
This allows us to calculate the average delay time $\qav{\delta \tau }$ for an electron 
to hop after the threshold ${\rm e}/2$ has been crossed by the charge  $Q(t)$:
\beq
	\qav{\delta \tau } = \int_0^{\infty}\!\!\!\!\!\! dt\; t\; {\cal P}(t)
	\,.
\eeq
In particular when $P_n$ can be approximated by 
the Gaussian \refe{sigmaGaussian} one obtains
$
\qav{\delta \tau }/\tau_s= \sqrt{\pi/(\kappa \rho)}
$
with $\qav{\delta \tau } \ll t_\star$.
To obtain the period of the SETOs one has to 
average the nonlinear expression \refe{Fgoth}:
$
	\mathcal{T}=\qav{\Fg}
$, which again in the SETOs region simplifies to
\beq
	\mathcal{T}=t_\star + 2\,\qav{\delta \tau }/(2+\kappa) \,.
	\label{SETOperiod}
\eeq

\begin{figure}\centering
\includegraphics[width=.45\textwidth]{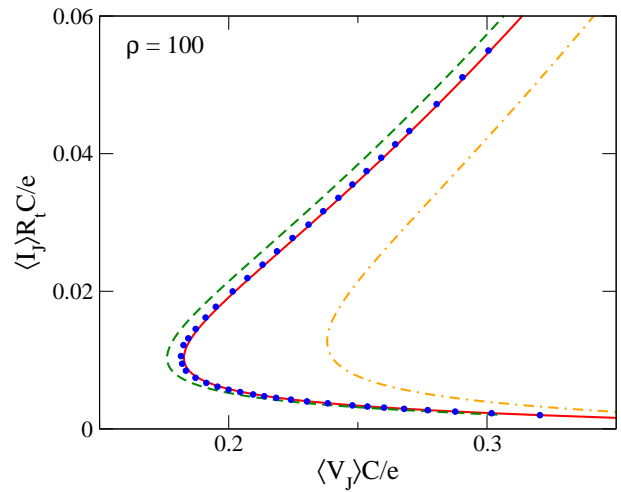}
\caption{\label{figNose} 
(Color online) Comparison between the different approximations for the 
SETOs ``nose'' of the current-voltage characteristics: Monte Carlo data (circles) are shown against 
analytical calculations from \refE{AppQ} (solid), \refE{fullQ} (dashed), and \refE{simpleQ} (dot-dashed).
}
\end{figure}

Let us now come back to the probability. 
Conservation of the probability gives that
$P_{n+1}(t)=1-P_n(t)$.
Since in this approximation the behavior is quasiperiodic,
the charge on the capacitor for the $n+1$ electron is 
{\em on average} $Q_f({\rm e}/2,t-t_n-\mathcal{T})$.
(A more precise discussion on the validity of this last average can
be found in the Appendix, where the problem is analyzed more 
rigorously.) 
The average charge can then be computed 
by averaging over a period as follows:
\begin{align}
\langle Q \rangle = \int_{t_n}^{t_n+\mathcal{T}}\!\! \frac{dt}{\mathcal{T}} \; \bigg[& Q_f\left({{\rm e}\over2},t-t_n\right) P_n(t) + \\ \nonumber
                 +\;& Q_f\left({{\rm e}\over2},t-t_n-\mathcal{T}\right) P_{n+1}(t)\bigg] \,.
\end{align}
In the limit of $\rho \gg 1$ only the gaussian part of the probability 
is relevant, and the integral gives:
\begin{align}\label{fullQ}
   \langle Q \rangle =&
   \Qs -\left(\Qs-{{\rm e}\over2}\right)\left(e^{\mathcal{T} \over \tau_s}-1\right){\tau_s\over \mathcal{T}}\; \times \\ \nonumber
   &\times\,\left[
   1-\sqrt{ \pi \over  \kappa \rho}\;
   e^{1 \over \kappa \rho} \; {\rm Erfc}\left({1/\sqrt{\kappa \rho}}\right)
   \right]
   \,.
\end{align}
\refE{fullQ} leads to $\langle V_J \rangle_{III}=C \langle Q \rangle$.
With little loss in the accuracy \refE{fullQ} can be simplified to the form:
\beq
	\langle Q\rangle \approx {{\rm e}\over 2} 
	\left[\kappa +1 -2 \left(1-\sqrt{ {\pi \over \kappa \rho}}\right) / \ln {\left(2+ \kappa \over \kappa\right)}\right]
	\,.
	\label{simpleQ}
\eeq
The analytical expressions \refe{fullQ}, \refe{simpleQ}, and \refe{AppQ} obtained 
in the appendix, are compared to the Monte Carlo results in 
\reff{figNose}.
%
%
%
%%%%%%%%%%%%%%%%%%%%%%%%%%%%%%%%%%%%%%%%
%
%%%%%%%%%%%%%%%%%%%%%%%%%%%%%%%%%%%%%%%%%
%
%
%
These expressions describe the current with good accuracy, and in particular they
all capture the presence of a minimum in the voltage $V_J$. 
This minimum signals the crossover region between two different 
kinds of SETOs.
We indicate them in \reff{regimes:fig} as III.1 and III.2.
The latter appears for $1 \ll \kappa \ll \rho$.
In this case, and in the extreme limit $\rho\rightarrow \infty$ the 
SETOs period becomes $\mathcal{T}={\rm e}/I_b$, i.e.\ corresponds exactly 
to the time needed to the ideal current source to furnish a charge ${\rm e}$.
The saturation value for the charge ($Q_s$) in this regime 
is much larger than ${\rm e}/2$, implying that only the linear
part of the exponential in \refE{Qfunction} is explored. 
This is important since the small fluctuations in the 
hopping times do not affect the evolution equation 
for the next electron.
One can readily verify that in the limit of an
ideal current source the charge time dependence 
around each $t_n$ is 
$Q(t)=I_b (t- n {\rm e}/I_b)$.
The non-linear corrections instead add a 
stochastic dependence on the time evolution
of the charge.
The period, for instance, does not depend 
on $\qav{\delta \tau}$ anymore for $\kappa \rightarrow \infty$,
as is clear from \refE{SETOperiod}.
This extreme limit is not realistic and thus the second term 
given in \refe{SETOperiod} is normally important.
%
%%%%%%%%%%%%%%%%%%%%%%%%%%%%%%%%%%%%%%%%
%
%
%             FIGURE 6
%
%
%%%%%%%%%%%%%%%%%%%%%%%%%%%%%%%%%%%%%%%%
\begin{figure}\centering
\includegraphics[width=.45\textwidth]{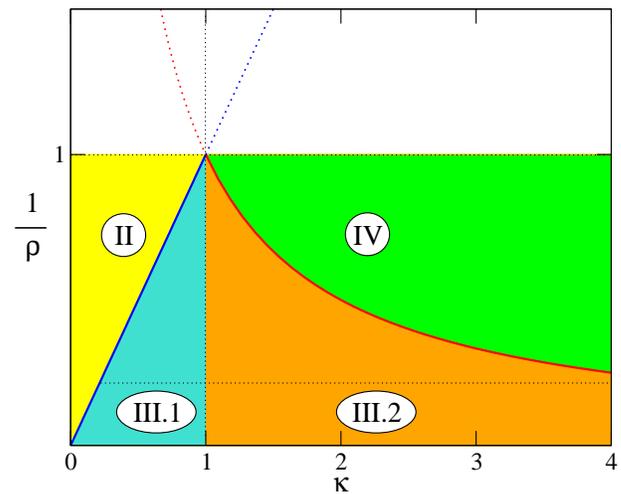}
\caption{
\label{regimescheme:fig} 
(Color online) Scheme of the boundaries of the transport regimes as a function of the relevant 
parameters: The tunneling/shunt resistances ratio $1/\rho=R_t/R_s$ and the 
relative distance of the current bias from the threshold $\kappa=(I_{b}-I_{\rm th})/I_{\rm th}$. 
The SETOs regime exists only in the limit $\rho \gg 1$.
}
\end{figure}
%%%%%%%%%%%%%%%%%%%%%%%%%%%%%%%%%%%%%%%

Reducing the current bias, the saturation charge $Q_s$ becomes of the order of 
${\rm e}/2$ ($\kappa=2Q_s/{\rm e}-1 < 1$) and the non-linear behavior of $Q(t)$ 
begins to correlate different hopping events. 
The charge time evolution in this regime is characteristic and 
resembles a shark fin, as shown in \reff{visualcharges:fig}.
It is also clear from \refe{SETOperiod}, that in this regime the stochastic
fluctuations have the greatest 
impact on the average SETOs period.

These two regimes can be identified on the current plot \reff{figNose} 
and in the analytical expression \refe{simpleQ} as the two branches
joined by a minimum of the voltage.
The large bias behavior ($\kappa\gg 1$) of $\refe{simpleQ}$ gives ${\rm e}\,\sqrt{\pi\kappa/\rho}\,/(2C)$, 
which is the Averin-Likharev expression for the current \refe{SETOextreme} ($\langle I_J\rangle \sim I_b$ in this limit), 
while in the opposite limit the 
long exponential charging time is dominating: 
$\qav{V_J}\approx ({\rm e}/C)\sqrt{\pi}/(\sqrt{\kappa \rho} \ln{(2/\kappa)})$.

The overall situation is summarized in \reff{regimescheme:fig}.
Since it is very difficult in practice to experimentally reach large values 
of $\rho$, the plot in \reff{regimescheme:fig} suggests
that a good experimental choice can be $\kappa=1$,
for which the SETOs appear for the lowest values of 
$\rho$.
We will discuss in the following the correlation function of the charge 
in order to analyze quantitatively the evolution of the accuracy of the SETOs.
%

%%%%%%%%%%%%%%%%%%%%%%%%%%%%%%%%%%%%
% SECTION III     CHARGE NOISE
%%%%%%%%%%%%%%%%%%%%%%%%%%%%%%%%%%%%

\section{Charge-fluctuation spectrum}
\label{SQQ}

A quantitative measure of the accuracy of the SETOs is given by 
the time correlation of the charge. 
This can be defined as 
\beq
	S(\tau) = \qav{Q(t+\tau) Q(t)} -\qav{Q(t+\tau)}\qav{Q(t)}
	\,,
	\label{Sdef}
\eeq
where the average is performed over a statistical ensemble 
and the result does not depend on $t$, since the stochastic 
process is stationary.
Note also that this quantity is proportional to the spectrum
of current fluctuations through the shunt resistance $R_s$:
$S_{R}(\tau)=S(\tau)/\tau_s^2$.
In the case of voltage-biased junction (see \reff{Model:fig}), 
$S_{R}$ gives the current fluctuations that can be directly 
measured through the load resistance $R$.

For perfectly periodic charge oscillations  
the Fourier transform of \refe{Sdef} is given by 
a sum of Dirac delta functions at $\omega=2\pi n /\mathcal{T}$, 
with $n$ integer. 
The non-periodic fluctuations introduce a finite width of these peaks.
The form of $S(\omega)$ measures thus directly and with a simple 
procedure the accuracy of the periodic charge transfer. 
In this section we derive an analytical expression for $S(\omega)$
that allows us to better understand the origin of the fluctuations.
In the next section we will compare these results to 
those obtained numerically by Monte Carlo simulations.

In order to calculate the Fourier transform of $S(\tau)$ for 
this stationary process it is
convenient to define the charge $Q(t)$ over a time $0<t<\Lambda$,
with $\Lambda \gg \mathcal{T}$ so that many SETOs are present in a single 
sample of $Q(t)$.
One can then calculate the Fourier series
\beq
	Q_p = \int_0^{\Lambda} {dt\over \Lambda} Q(t) e^{i p {2\pi\over \Lambda} t }
	\quad , \quad
	Q(t)=\sum_p e^{-i p  {2\pi  \over\Lambda} t } Q_p
	\label{FourierSeries}
	\,.
\eeq
Substituting \refE{FourierSeries} into \refE{Sdef} and averaging over 
$t$ one obtains:
\beq
	S(\tau)=\sum_{p} \qav{|Q_p|^2} e^{ - i p {2 \pi \over \Lambda} \tau}
	-Q_0^2
	\,,
\eeq
which can be used numerically to compute the correlation function from the 
Monte Carlo data, or analytically, by performing the limit $\Lambda \rightarrow \infty$.
In particular the Fourier transform can be defined as  
\beq
	S(\omega) = \int_{-\Lambda/2}^{+\Lambda/2} d\tau e^{i\omega \tau -0^+ |\tau|} S(\tau)
	\,,
\eeq
which gives 
\beq
	\label{Somega}
	S(\omega) = \sum_{p} (\qav{|Q_p|^2}-\delta_{p,0}Q_0^2) 2 \pi \delta(\omega- \omega_p)
	\,,
\eeq
with $\omega_p = 2 \pi p /\Lambda$. 
The presence of the Dirac delta functions is an artifact due to the periodic 
extension induced by the Fourier transform. 
In practice, since the frequency scale $1/\Lambda$ is infinitesimal one can
obtain the smooth function $S(\omega)$ by averaging the expression
\refe{Somega} for each value of $\omega$ over a small interval $2 \pi/\Lambda$.
This simply gives that
\beq\label{Swinkhinc}
	S(\omega_p) = \Lambda \qav{|Q_p|^2}
\eeq
for $p\neq 0$ (Wiener-Khinchin theorem).

The problem is now reduced to the calculation of the Fourier series of the charge.
Using the definitions of $t_n$ and $\tau_n$ given before in \refE{xi} and assuming 
that the extrema of the time evolution of $Q(t)$ coincide with the two hopping 
events at times $\tau_0$ and $\tau_N$ we can write:
\begin{align}
  Q_p
  =&
  \sum_{n=0}^{N-1}
  \int_{\tau_n}^{\tau_{n+1}}
    { dt \over \Lambda} 
    \,Q(t)\,e^{i\omega_p t} =
  \\ \nonumber
  =& 
  \sum_{n=0}^{N-1} 
  {e^{i\omega_p\tau_n}}\!\!
  \int_0^{\tau_{n+1}-\tau_n}
    \!\!
    { dt  \over \Lambda} \;
    Q_f({\rm e}/2, t-t_{n+1}+\tau_n) \,e^{i\omega_p t}\,.
\end{align}
In the limit of well-established SETOs the integral gives a contribution
that fluctuates very little. On the contrary the exponentials are much more 
sensitive to even small fluctuations of the tunneling times, since the phase 
results from the accumulation of many different hopping events. 
For this reason we expect that the upper integration limit can 
be substituted with the period of the SETOs $\tau_{n+1}-\tau_{n} 
\approx \mathcal{T}$ and we use $Q_f({\rm e}/2,t-(\mathcal{T}-\qav{\delta \tau}))$ as 
the average charge dependence.
The Fourier transform then takes the form:
\beq
	\label{finalSom}
	S(\omega)  = N \mathcal{T}  \qav{|F(\omega)|^2}  {\cal A}(\omega)
	\,,
\eeq
where
\begin{equation}
	\label{Fdef}
	F(\omega) 
	= 
	{1 \over N} \sum_{n=0}^{N-1} e^{ i \omega t_n}
\end{equation}
and
\beq
	\label{Adef}
	{\cal A}(\omega) = 
	\left|e^{i \omega \qav{\delta \tau}} \!\!
	\int_0^{\mathcal{T}}  \!\!
	{dt\over \mathcal{T}} \, Q_f({\rm e}/2,t-\mathcal{T}+\qav{\delta \tau})
	e^{i \omega t}
	\right|^2
	\,.
\eeq
The quantity ${\cal A}$ can be readily evaluated:
\beq\label{Aexplicit}
\begin{split}
	{\cal A}&(\omega)= \left({{\rm e}\tau_s\over 2 \mathcal{T} }\right)^2 \times \\
	\times & \left| 
	{\kappa +1 \over \omega \tau_s}(e^{ i\omega \mathcal{T}}-1)
	-{\kappa e^{-\qav{\delta \tau}/\tau_s} \over \omega \tau_s +i}(e^{i \omega \mathcal{T}}-e^{ \mathcal{T}/\tau_s})
	\right|^2
	\,.
\end{split}
\eeq

In order to proceed we have to evaluate also the average 
of $F(\omega)$. 
It is convenient to express the time at which one event happens 
as a sum over the delays between previous events
using \refE{Fgothdef}:
\beq
	F(\omega)
	={e^{i \omega t_0} \over N} 
	\left( 
	1+
	\sum_{n=1}^{N-1} 
	\exp\left\{ i \omega \sum_{k=0}^{n-1}  \mathfrak{F}(\delta \tau_k)\right\}
	\right) \,.
\eeq
Now the average of $|F(\omega)|^2$  can be performed using the 
distribution function ${\cal P}(\delta \tau)$:
\beq
	\qav{|F(\omega)|^2}
	=
	{1\over N} 
	\left(1+2 {\rm Re} \left\{ {g(\omega) \over 1-g(\omega)}\right\}\right)
	+{\delta F\over N^2}
	\label{Fform}
\eeq
where we introduce the quantities:
\beq
	g(\omega) =\qav{e^{i \omega \mathfrak{F}(\delta \tau)}}
	 = 
	\int_0^\infty \!\!\!\!\!\! d (\delta \tau) {\cal P}(\delta \tau) e^{i \omega \mathfrak{F}(\delta \tau)}
	\label{gdef}
\eeq
and  
$\delta F=2 {\rm Re}\left\{ g(g^{N}-1) /(1-g)^2 \right\}$,
whose contribution to $S(\omega)$ vanishes in the limit $N\rightarrow \infty$.
In conclusion we find:
\beq\label{SomNinfinity}
	S(\omega) = \left(1+2 {\rm Re} \left\{ {g(\omega) \over 1-g(\omega)}\right\}\right)
				\mathcal{T} {\cal A}(\omega)\,,
\eeq
which constitutes the central result of this section.

We are now in the position to study the spectrum of the charge fluctuations 
for the system at hand. 
As it can be seen from the form of \refe{Fform} the function has a singularity
for $g\rightarrow 1$. 
Since 
\beq
	|g(\omega)|^2 = \int_0^\infty \!\!\!\!\! dt \int_0^t \!\!\! dt'\; {\cal P}(t) {\cal P}(t') 
	2 \cos( \omega t) \leq 1   
\, ,
\eeq
 a singularity is present for $\omega \rightarrow 0$ or 
when the fluctuations are negligible so that 
$g\approx e^{i \omega \qav{\mathfrak{F}(\delta \tau)}}$.
This picture predicts a series of peaks for the  
frequencies $\Omega_n = 2 \pi n/\mathcal{T}$ with $n$ integer.
Small fluctuations introduce a finite width, regularizing 
the correlation function. 
The numerical integration in the expression of  $g(\omega)$ is straightforward, 
but it is also possible to obtain an analytical expression.
Deep in the SETOs regime ${\cal P}(\delta \tau)$ has a gaussian 
behavior and provides a short cutoff time, so that 
we can expand the exponential in \refe{gdef} to second order in 
$\delta \tau$.
This gives 
\begin{align}\label{gexpand}
  g(\omega)& = e^{i \omega t_\star} \; \times \\ \nonumber
  \times &\left ( 1 + {2i \omega \over 2+\kappa} \qav{ \delta \tau}
  - {\kappa^2 i \omega+4 \omega^2\tau_s 
  \over 2(2+\kappa)^2 } {\qav{\delta \tau^2}\over \tau_s}
  + \dots \right)
  \,.
\end{align}

The maximum of ${\rm Re}\{g/(1-g)\}$ takes place 
for ${\rm Arg}[g(\omega)]=0$ that at lowest order  in $\qav{\delta \tau}$ 
gives for the position of the poles:
\beq
	\Omega_n 
	= 
	{2 \pi n \over t_\star}
	\left(1- {2\qav{\delta \tau} \over (2+  \kappa)t_\star }\right) 
	\,,
\eeq
coinciding at linear order in $\qav{\delta \tau}/\tau_s$ 
with ${2 \pi n/\mathcal{T}}$.
The phase of $g$ ($g=|g|e^{i \phi}$)
thus vanishes at the minimum of ${\rm Re}\{g/(1-g)\}$, so that near this point 
one can write at lowest order $\phi \approx \mathcal{T} (\omega-\Omega_n)$.
The relevant part of \refe{Fform} then reads:
\beq\label{Reg1sug}
	{\rm Re}\left\{g \over 1-g\right\}\approx {(1-|g|) \over  (1-|g|)^2+  \phi^2 }
	\,,
\eeq
and \refE{finalSom} takes the simple Lorentzian form 
\beq
S(\omega)\simeq\mathcal{A}(\Omega_n)\,
{\Gamma_n/2 \over \Gamma_n^2/4+(\omega-\Omega_n)^2}
\label{SofOmega}
\,,
\eeq
with the full width at half maximum $\Gamma_n$ defined by:
\beq\label{widthgam}
  \Gamma_n=2\;{1-|g| \over \mathcal{T}}={4\Omega_n^2 \over \mathcal{T} (2 + \kappa)^2} \left(\qav{\delta \tau^2}-\qav{\delta \tau}^2\right)\,.
\eeq
The presence in this formula of the mean squared variance of 
delay in the tunneling time, $\qav{\;(\delta \tau-\qav{\delta \tau})^2}$,
clearly indicates that the spread in the hopping events controls the width
of the peak, as is physically expected. 
One also sees that the width of the poles increases 
with $n$.
Performing the average with the gaussian distribution 
we find $\qav{\;(\delta \tau-\qav{\delta \tau})^2}/\tau_s^2=  4(1-\pi/4)/(\kappa\rho)$
and for the relative width $\Gamma_n/\Omega_n$ we have the explicit expression:
\beq
	\label{DeltaOmega}
	{\Gamma_n \over \Omega_n} = 
	{32 \pi \, n \, (1-\pi/4) \over  \rho \, \kappa \, (2+\kappa)^2 \, \ln^2\left[(2+\kappa)/\kappa\right]}
	\,.
\eeq
\begin{figure}\centering
\includegraphics[width=.45\textwidth]{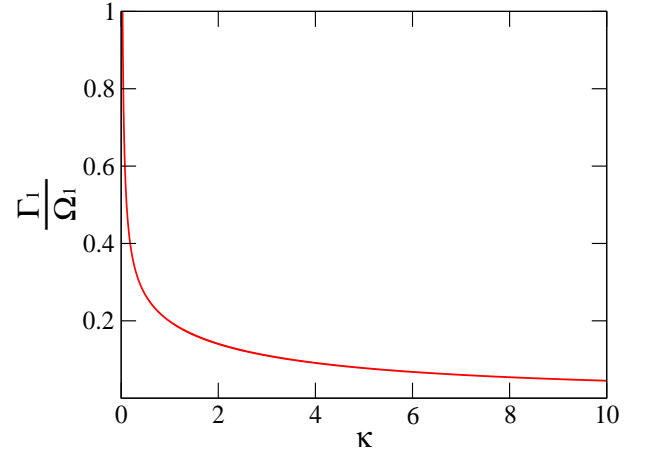}
\caption{\label{spreadk} 
The relative width at half height of the first noise peak as a function of $\kappa$, given by \refE{DeltaOmega}, for $\rho=10$. 
}
\end{figure}
%%%%%%%%%%%%%%%%%%%%%%%%%%%%%%%%%%%%%%%%
%
%
%

\refE{DeltaOmega} allows to study the width of the peak in the 
charge-fluctuation spectrum, and thus, the accuracy of the SETOs.
It correctly gives that $\Gamma_1 / (2\pi/\mathcal{T}) \ll 1$ in the SETOs regime 
($ 1/\rho \ll\kappa \ll \rho$).
It also shows (see \reff{spreadk}) that the \textit{relative} width $\Gamma_1/\Omega_1$ is a 
monotonic decreasing function of the bias current ($\kappa$) for a given value of the resistance ($\rho$).

From \refE{SofOmega} it is clear that within this approximation
the weight of the Lorentzian peak is controlled  only by the 
form factor ${\cal A}(\Omega_n)$.
Using its explicit expression one finds that for large $\rho$ it 
becomes independent of $\rho$: ${\cal A}(\Omega_1) \simeq 
{\rm e}^2/\left(4\pi^2 + \ln^2\left[\frac{2+\kappa}{\kappa}\right]\right)$.
The full variation takes place in the $\kappa<1$ region, where 
the shape of $Q(t)$ evolves from the shark-fin to the sawtooth form.
From the form of \refE{DeltaOmega} one can also see that 
for small $\kappa$ the peak broadens and the SETOs are washed 
out when $\Gamma_1 \sim \Omega_1$ (for $\kappa \sim 1/\rho$).
For large $\kappa$ the theory instead predicts that the relative 
width decreases monotonically \footnote{
This expression does not agree with the expression (57)  of 
Ref.~\cite{AverinLikharev}, specifically we find a different 
functional dependence on $R_s$: $\Gamma \sim 1/R^2_s$ instead 
of $1/R_s$.
}:
\beq
	\label{DeltaOmega2}
	{\Gamma_n \over \Omega_n} =  
	{8 \pi n (1-\pi/4) \over  \rho \kappa }
	\,.
\eeq
In this limit the SETOs disappear by a decrease of the weight
of the peak, but within our approximation this is not seen.
Actually for sufficiently large current bias ($\kappa \sim \rho$) 
there is a finite probability that a single tunnel event is no more 
sufficient to bring the charge back in the Coulomb blockade region.
This is quantified by the value of  $P_n(t_n)$, which,  
contrary to our hypothesis, can become smaller than $1$.
One can expect the theory to roughly remain valid for 
the fraction of tunneling events that leaves $Q(t_n)<{\rm e}/2$.
This describes a peak that remains sharp, but that vanishes 
in weight as $P_n(t_n)$. 

Another interesting and relevant limit is the low-frequency behavior 
of $S(\omega)$. 
Expanding \refE{gdef} in $\omega$ one can show that
\begin{align}
	N\qav{|F(\omega)|^2}
	= \;& {F_2-F_1^2 \over F_1^2}\; + \\ \nonumber 
	+\; & \omega^2 \; {4F_1 F_2 F_3-F_1^2 F_4 - 3 F_2^3 \over 12 F_1^4}
	+ \dots
\end{align}
where $F_n=\qav{\mathfrak{F}^n}$. 
If the fluctuations are negligible then $F_n=F_1^n$ and 
the noise at low frequency vanishes.
In particular in our case this average takes a simple form if the 
explicit expression of $\mathfrak{F}$ is used:
\beq\label{zerofreqfact}
	N\qav{|F(\omega)|^2}
	\simeq 4 {\qav{\delta\tau^2}-\qav{\delta \tau}^2 \over t_\star^2 (2+\kappa)^2 }	
	\left(1+{\omega^2 t_\star^2 \over 12}+\dots\right)
	\,.
\eeq
We thus find that in the SETOs regime the low-frequency noise is suppressed.
%%%%%%%%%%%%%%%%%%%%%%%%%%%%%%%%%%%%%%%%%%%%%%%%%%%%%%%%%%%%%
% FANO FACTOR DISCUSSION
Eq.~\eqref{zerofreqfact} together with the expansion of Eq.~\eqref{Aexplicit} for $\omega\rightarrow 0$ 
allows us to evaluate the zero-frequency Fano factor:
$\mathbb{F}\equiv {S_J(0)}/{{\rm e}\qav{I_J}}=S(0)/{\tau_s^2{\rm e}\qav{I_J}}$, where $S_J(\omega)$ is the 
noise spectrum of the current through the junction and the relation
$S(\omega)=S_J(\omega)\tau_s^2/(1+\omega^2\tau_s^2)$ holds exactly.
The reduction of the current fluctuations in the large-$\rho$ limit 
naturally leads to sub-Poissonian noise ($\mathbb{F}<1$)
vanishing for $\rho\rightarrow \infty$:
\begin{equation}
	\mathbb{F}
	\simeq
	\frac{(4 - \pi)(-2 + (\kappa+1)t_\star/\tau_s)^2}
				{\rho\,\kappa \,(2 + \kappa)^2  (t_\star/\tau_s)^2}\,.
\end{equation}
Nevertheless this expression is only qualitatively correct, 
since the analytical theory
has been designed to describe accurately the noise for frequencies 
around the peak of the SETOs.
This will be shown in the next section, where we present numerical simulations. 
%
%%%%%%%%%%%%%%%%%%%%%%%%%%%%%%%%%%%%%%%%%%%%%%%%%%%%%%%%%%%%%%
%

%%%%%%%%%%%%%%%%%%%%%%%%%%%%%%%%%%%%
% NUMERICAL SIMULATIONS
%%%%%%%%%%%%%%%%%%%%%%%%%%%%%%%%%%%%

\section{Numerical simulations}
\label{MCresults}

In this section we show numerical results obtained by Monte Carlo 
simulations for the charge-fluctuation spectrum.
The purpose of this section is to compare these results with the 
analytical calculations of the preceding sections, valid for $\rho\gg1$ and 
$1/\rho \ll \kappa \ll \rho$, and to explore the crossover region
where the SETOs disappear.

% COMMENTI TECNICI SUL MONTE CARLO

The Monte Carlo simulations are performed by generating
different realizations of the stochastic time evolution of the charge $Q(t)$
over a time much longer than the SETOs period.
The time evolution of the charge is obtained by discretizing the time 
on a nonuniform grid, such that in the time interval $\Delta t$ the 
charge varies by a small quantity and 
$P=\Gamma(Q(t)) \Delta t < \mathcal{N} \ll 1 $.
The tunneling event is accepted or refused 
by generating a random number between 0 and 1 
and by comparing it to $P$. 
The deterministic evolution of $Q(t)$ between two events 
is simply given by \refE{Qfunction}.
The sequence of time intervals of deterministic evolution interspersed by 
tunneling times so constructed gives the full knowledge of $Q(t)$ and 
constitutes the stochastic run. 
The square modulus of the Fourier transform of the charge is then 
easily calculated analytically piecewise, interval by interval. 
To obtain the noise as from \refe{Swinkhinc}, 
just a further average over several runs is needed.
Typically $\mathcal{N}=.01$, each run counts $10^3$ tunneling 
events, and an average over $10^4$ realizations is performed.

\begin{figure}
\includegraphics[width=.45\textwidth]{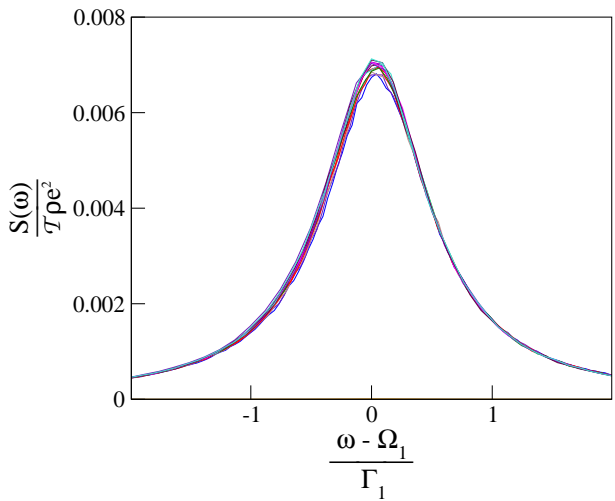}
\caption{\label{noisewidthcompare} 
(Color online) The noise peak at $\omega=\Omega_1$ obtained from Monte Carlo numerical simulations for values of $\rho$ ranging from $80$ to $200$ 
in steps of $10$. On the $x$-axis the frequency is shifted by $\Omega_1$ and scaled by the width 
$\Gamma_1$ as given by \refE{widthgam}, on the $y$-axis the noise is scaled by $\mathcal{T}\rho$.
}
\end{figure}

Let us now discuss the numerical results.
We begin by comparing the form of the first peak in the 
noise spectrum. 
It is shown scaled by the analytically calculated width 
$\Gamma_1$ in \reff{noisewidthcompare} for different 
values of $\rho$ and given $\kappa$. 
The agreement is excellent.
%
%%%%%%%%%%%%%%%%%%%%%%%%%%%%%%%%%%%%%
%
%             FIGURE 9
%
%%%%%%%%%%%%%%%%%%%%%%%%%%%%%%%%%%%%%
%
%%%%%%%%%%%%%%%%%%%%%%%%%%%%%%%%%%%%%%%
%
We then compare the full $\omega$ dependence of $S(\omega)$ obtained 
from \refE{SomNinfinity} with the one calculated numerically.
%
%%%%%%%%%%%%%%%%%%%%%%%%%%%%%%%%%%%%%%%%
%
%          FIGURE 10
%
%%%%%%%%%%%%%%%%%%%%%%%%%%%%%%%%%%%%%%%
\begin{figure}
\includegraphics[width=.45\textwidth]{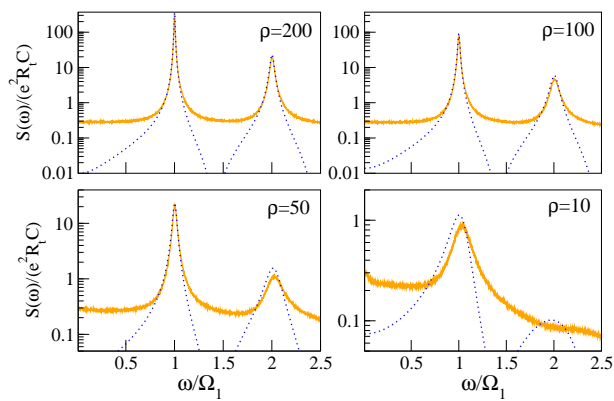}
\caption{\label{noisecompare} 
(Color online) Comparison between Monte Carlo data for the charge-fluctuation spectrum and the analytical results obtained from \refE{SomNinfinity} 
for different values of $\rho$, at fixed bias condition $\kappa=1$.
}
\end{figure}
%%%%%%%%%%%%%%%%%%%%%%%%%%%%%%%%%%%%%%%
%
We show the comparison in \reff{noisecompare} for the relevant case 
$\kappa=1$ (which corresponds to the widest extension in $\rho$ of the SETOs region) %, as can be seen from \reff{regimescheme:fig}) 
and for $\rho$ ranging from $200$ to $10$. 
As expected, the agreement is very good deep in the SETOs regime, for $\rho\gg1$, and at $\rho=10$ all the essential features of the peak 
at $\omega=\Omega_1$ are still fairly well represented by the theory. 
The analytical approximation correctly finds the main contribution to the 
noise, but it tends to underestimate it far from the peak maximum.

%%%%%%%%%%%%%%%%%%%%%%%%%%%%%%%%%%%%%%%%
%
%  FIGURE 11
%
%%%%%%%%%%%%%%%%%%%%%%%%%%%%%%%%%%%%%
\begin{figure}
\includegraphics[width=.45\textwidth]{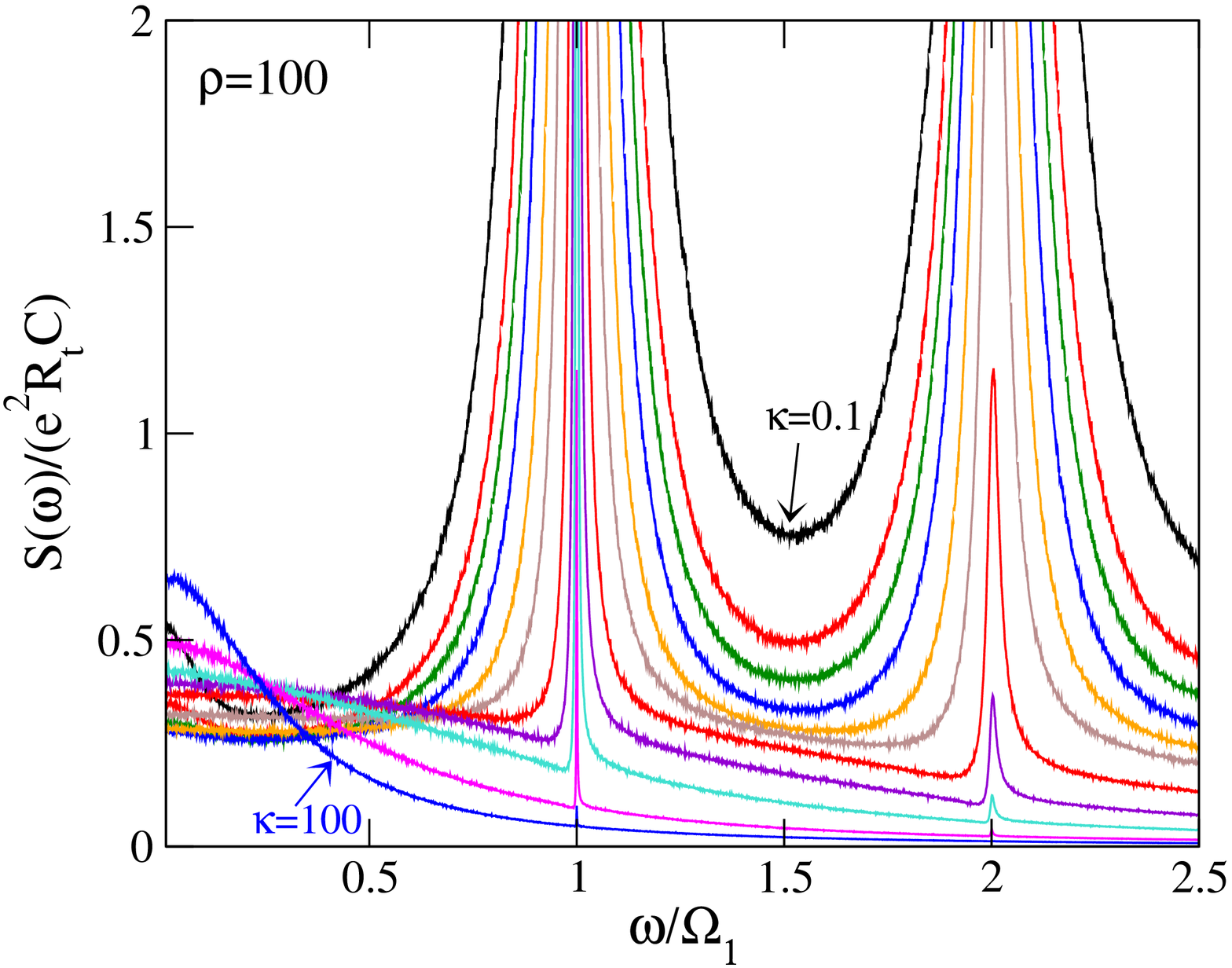}\\
\includegraphics[width=.45\textwidth]{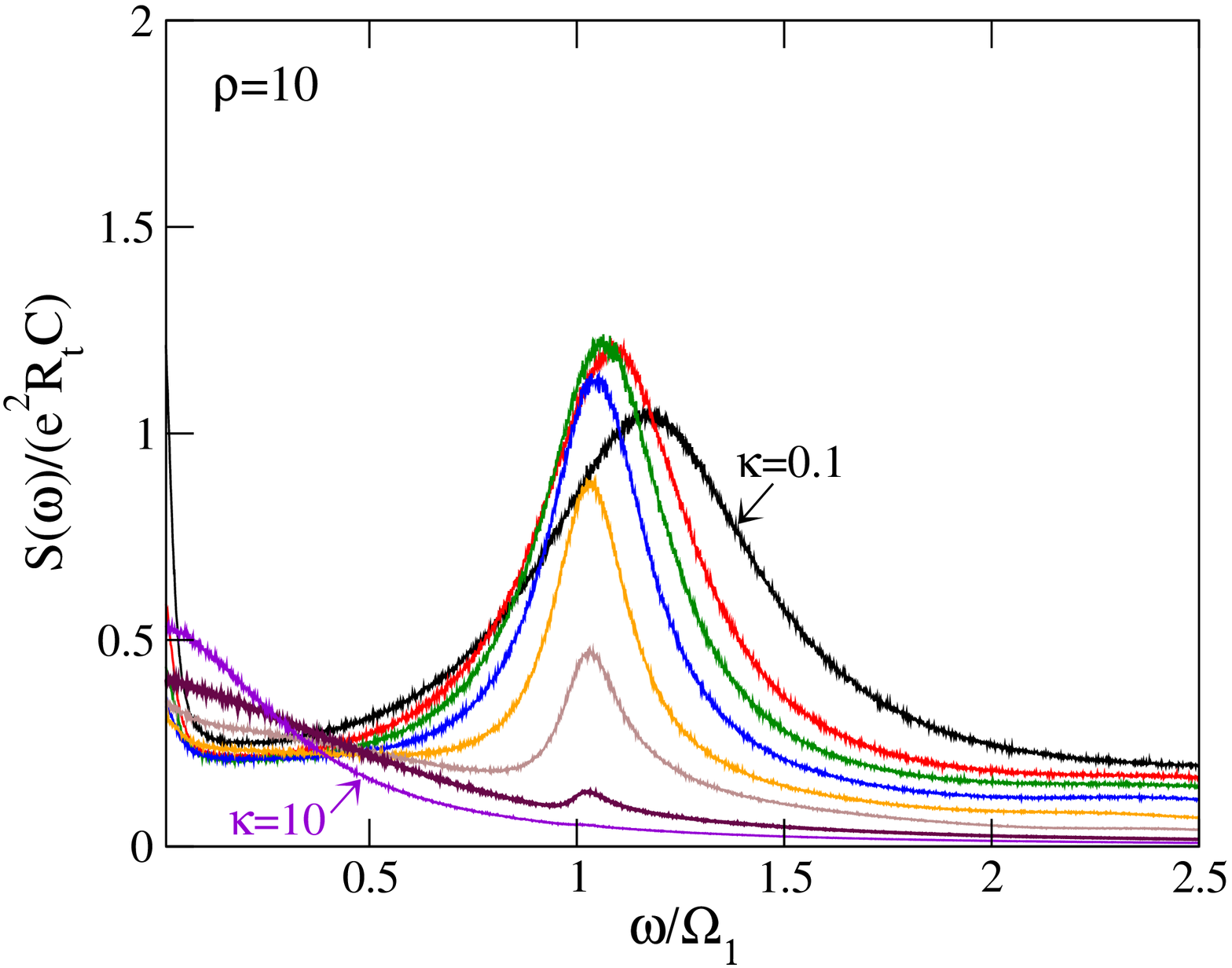}
\caption{\label{noisekvary} 
(Color online) Monte Carlo spectra for different bias conditions $\kappa$ at fixed junction environment: $\rho=100$ ($\rho=10$) in the upper (lower) graph.
Increasing $\kappa$ means here moving along a horizontal line in \reff{regimescheme:fig} toward the high-bias boundary of the SETOs region 
and it allows us to see how SETOs disappear.
}
\end{figure}
%%%%%%%%%%%%%%%%%%%%%%%%%%%%%%%%%%%%%%%
%

Let us now investigate how the SETOs disappear. 
From the experimental point of view a simple parameter that can be varied continuously
is $\kappa\sim I_b$. 
We thus plot in \reff{noisekvary} the 
evolution of $S(\omega)$ for given $\rho=10$ and 100 as a function of 
$\kappa$.
These two plots show several interesting features.
The first striking one is the reduction of the relative widths of the 
peaks by increasing $\kappa$. 
This is predicted by the analytical expression \eqref{DeltaOmega} and 
the numerical calculations assess its validity even outside the region of 
applicability of the analytical theory.
Note that in \reff{noisekvary} the frequency axis is rescaled with 
$\Omega_1$, thus the apparent weight of the peaks is reduced by 
the scaling, but it saturates in the region $\kappa \ll \rho$
as predicted by the analytical theory, and then starts to decrease in the crossover region.
The second visible feature is the appearance of a wide Lorentzian
zero frequency peak that remains the only structure for $\kappa \gg \rho$.
This structure is due to the charge noise induced at the capacitance by the 
Poissonian current fluctuations generated by the tunnel junction.
By solving the electromagnetic problem one finds: 
$S(\omega) = C^2 |Z(\omega)|^2 e \qav{I_J}$,
where $e \qav{I_J}$ is the standard 
tunnel junction Poissonian white noise and 
$Z(\omega)=R_\parallel/(1+ i \omega R_\parallel C )$,
with $R_\parallel=R_s R_t/(R_s+R_t)$,
the impedance between the current source and the voltage at the 
capacitance.
This gives:
\beq
        S(\omega)
	=
	{{\rm e}\langle I_J\rangle C^2R_s^2 \over 1+2 \rho +
	\rho^2(1+     \omega^2R_t^2C^2)}\,,
\eeq  
which fits our data very well (not shown).
%%%%%%%%%%%%%%%%%%%%%%%%%%%%%%%%%%%%%%%
%
%        FIGURE 12   RHO Dependence 
%
%%%%%%%%%%%%%%%%%%%%%%%%%%%%%%%%%%%%%%%
\begin{figure}
\includegraphics[width=.45\textwidth]{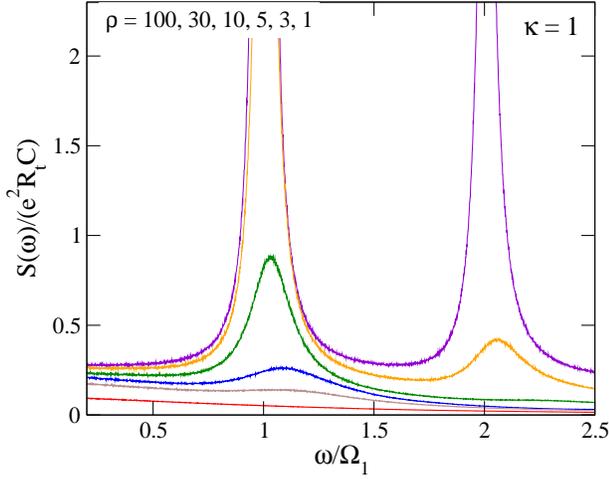}
\caption{\label{noiserhovary} 
(Color online) Monte Carlo spectra for fixed bias conditions $\kappa=1$ and different values of $\rho$: The crossover from the SETOs regime to 
the $\rho\lesssim 1$ region is shown.
}
\end{figure}
%%%%%%%%%%%%%%%%%%%%%%%%%%%%%%%%%%%%%%%

Finally we show in \reff{noiserhovary} the evolution of $S(\omega)$
for $\kappa=1$ and for $\rho$ evolving from 1 to 100.
This figure gives an idea of the expected spectrum at the 
optimal value $\kappa=1$ as a function of $ \rho$. 
It turns out that already at $\rho=3$, $S(\omega)$ presents 
a very broad maximum and $\rho=5$ is probably sufficient to observe 
a clear structure in $S(\omega)$.

%%%%%%%%%%%%%%%%%%%%%%%%%%%%%%%%%%%%
% CONCLUSIONS
%%%%%%%%%%%%%%%%%%%%%%%%%%%%%%%%%%%%

\section{Conclusions}
\label{conclusions}

In this paper we have studied theoretically electronic 
transport in a tunnel junction in the presence of a large-resistive environment.
The phenomenon of SETOs has been predicted to appear 
in this system for essentially infinite value of the 
external resistance, so that the junction can be seen as 
current biased. 
We investigated under which conditions the SETOs 
appear for a realistic finite value of the environment 
resistance.
We found analytical expressions for the current
[Eq.~\refe{fullQ}] and for the 
charge-fluctuation spectrum [Eq.~\refe{finalSom} 
with Eq.~\refe{widthgam}].
Our analytical results describe very well the form of the 
peak in the charge noise, which can be regarded as the 
hallmark
of the SETOs, since it quantifies the 
accuracy of the periodicity in the charge time dependence.
We find that a ratio of $R_s/R_t$ as low as  
$5$ can be sufficient to observe a clear structure in 
$S(\omega)$ if the bias current is chosen such that 
$I_b\approx 2 I_{\rm th} = {\rm e}/R_s C$ (this can be 
converted on a condition on the voltage bias 
$V_b \approx {\rm e}/C$).
This ratio can be obtained experimentally and thus 
in principle SETOs can be observed through the measurement of 
the current or charge noise. 

The conclusion on the possibility of observing the effect is thus 
optimistic. 
One should however keep in mind that, 
following the literature on this problem, the theory presented 
holds at low temperature ($ k_B T \ll {\rm e}^2/2C$) and 
neglects quantum fluctuations of the electromagnetic 
modes of the environment.
Thermal and quantum fluctuations are expected to have a non-negligible influence 
on the transport mechanism at play in the system. 
Their evaluation requires a different technical 
approach and is beyond the scope of the present paper.

\section*{Acknowledgements}

We are indebted for useful correspondence with Y. Pashkin.
We also gratefully acknowledge fruitful discussions with D. Esteve, 
F. Portier, and  L.S. Kuzmin.
Comments on the manuscripts are acknowledged from R. Avriller, V. Puller, and M. Houzet. 
We finally acknowledge partial financial support through the French 
ANR grant QNM No.\ 0404 01.

%%%%%%%%%%%%%%%%%%%%%%%%%%%%%%%%%%%%
% APPENDIX
%%%%%%%%%%%%%%%%%%%%%%%%%%%%%%%%%%%%%

\appendix
\section{CALCULATION OF THE $I$-$V$ CHARACTERISTICS IN THE 
SETOs REGIME WITH THE MASTER EQUATION APPROACH}

\label{appA}

In this Appendix we find for the $I$-$V$ characteristics expressions 
which take into account the spread in the distribution probability 
of the charge.
A full statistical description of the behavior of the system can 
be given in terms of the probability $\sigma_n(Q,t)$ that at time $t$ 
the charge in the capacitance is $Q$ and $n$ charges have 
crossed the junction.
Conservation of the probability and the master equation describing 
electron tunneling lead to the following set of coupled 
partial differential equations:
\begin{eqnarray}\label{PartialEq}
	\lefteqn{\frac{\partial \sigma_n(Q,t)}{\partial t} =
	\frac{\partial}{\partial Q}\left[\frac{(Q-Q_s)}{\tau_s}\sigma_n(Q,t)\right]}&&\\ \nonumber
   	&&-\Gamma(Q/C) \sigma_n(Q,t)+ \Gamma((Q+{\rm e})/C) \sigma_{n-1}(Q+{\rm e},t)
	\,.
\end{eqnarray}
\refE{PartialEq} is a generalization of the equations 
given in Ref.~\cite{AverinLikharev} to include the information
on the number of electrons which have tunneled.
The general solution of this equation is difficult in the presence of 
a finite resistance.
But in the SETOs regime we can find an explicit solution by 
exploiting the fact that at every cycle the charge passes 
through the blocked range ($-{\rm e}/2 < Q < {\rm e}/2$). 
Let us assume that at $t=0$ the distribution function is:
\begin{equation}
	 \sigma_n(Q,0)=\delta(Q-{\rm e}/2)  \delta_{n,n_0}
	\,.
\end{equation}
The differential equation \refe{PartialEq} for $\sigma_{n_0}(Q,t)$ can 
be easily solved, since it decouples from the others ($\sigma_{n_0-1}=0$):
\beq
	\sigma_{n_0}(Q,t)=P(t)\delta\left(Q-Q_f\left(\frac{{\rm e}}{2},t\right)\right)\,,
\eeq
where $P(t)=P_n(t)$ as given by \refE{sigmaGaussian} with $t_n=0$ 
and $Q_f$ is defined in \refE{Qfunction}.
Once we know the solution for $\sigma_{n_0}$ we can substitute it into
\refE{PartialEq} and find the solution for $\sigma_{n_0+1}$.
This can be done by using the ansatz:
\beq
    \sigma_{n_0+1}(Q,t)=\int\!dQ^\prime
   \delta\left(Q-   Q_f\left(Q^\prime,t\right)\right)f(Q^\prime,t)\,,
\eeq
that gives 
\beq
       \sigma_{n_0+1}(Q,t)=f(z(Q),t)e^{\frac{t}{\tau_s}}\,,
\eeq
with $z(Q)=(Q-Q_s)e^{\frac{t}{\tau_s}}+Q_s$, and $f(z(Q),t)=0$ for $z(Q)<{\rm e}/2$ and $t>\tau_s \ln\left(\frac{1}{2}-\frac{z(Q)}{{\rm e}}\right)$.
The resulting differential equation for $f$ reads:
\begin{align}
 \frac{\partial f(z(Q),t)}{\partial t}&=\left(Q_s+\frac{{\rm e}}{2}+(z(Q)-Q_s)e^{-\frac{t}{\tau_s}}\right)\times \\ \nonumber
  \times& e^{-\frac{t}{\tau_s}} \frac{P(t)}{{\rm e}R_tC}\;\delta\left({\rm e}+\left(z(Q)-\frac{{\rm e}}{2}\right)e^{-\frac{t}{\tau_s}}\right)\,.
\end{align}
The equation can be integrated and gives for $\sigma_{n_0+1}(Q,t)$
\begin{align}
 \sigma_{n_0+1}(Q,t)=&e^{\frac{t}{\tau_s}}\,P\left(\tau_s\ln\left(\frac{1}{2}-\frac{z(Q)}{{\rm e}}\right)\right)\times \\ \nonumber
 \times & \frac{\tau_s}{{\rm e}R_tC}\frac{\left(\frac{{\rm e}}{2}-Q_s\right)\left(z(Q)+\frac{{\rm e}}{2}\right)}{\left(z(Q)-\frac{{\rm e}}{2}\right)^2}
\end{align}
that is non vanishing for $Q_1(t)<Q<Q_2(t)$ where
$Q_1(t)=Q_s-{\rm e}-(Q_s-{{\rm e}}/{2})e^{-{t}/{\tau_s}}$ 
and 
$Q_2(t)=Q_s-(Q_s+{{\rm e}}/{2})e^{-{t}/{\tau_s}}$.  
It is interesting to note that the distribution has now a finite 
spread in $Q$ induced by the combined action of the 
stochastic fluctuations and of the finite value of the resistance.
This is in contrast with the simpler approximation used in the text
to evaluate the average in \refE{fullQ}, where we assumed that 
the spread was negligible, and that a delta function could be used 
to describe the distribution $\sigma_{n_0+1}$.

We can now calculate the average charge on the junction during a single 
oscillation:
\begin{equation}\label{AppQ}
  \langle Q\rangle 
	=
	\int_0^{\mathcal{T}}\!\! \frac{dt}{\mathcal{T}} 
	\left(
	Q_f\left(\frac{{\rm e}}{2},t\right)P(t)+
	\int_{Q_1(t)}^{Q_2(t)} \!\!\!\!\!\!\!\!\!\!\! dQ \;Q\; \sigma_{n_0+1}(Q,t)
	\right) \,.
\end{equation}
The numerical integration of this expression leads to the result shown in 
\reff{figNose}.

\bibliographystyle{unsrt}
\bibliography{bibliomeso}

--------------------

\end{document}